\documentclass[letterpaper,twocolumn,american,showpacs,prl,aps,superscriptaddress]{revtex4}
\usepackage[latin1]{inputenc}
\usepackage{bm}
\usepackage{amsmath}
\usepackage{graphicx}
\usepackage{amssymb}

\begin{document}

\title{Quantum Semi-Markov Processes}

\author{Heinz-Peter Breuer}

\affiliation{Physikalisches Institut, Universit\"at Freiburg,
             Hermann-Herder-Strasse 3, D-79104 Freiburg, Germany}

\affiliation{Hanse-Wissenschaftskolleg, Institute for Advanced
             Study, D-27753 Delmenhorst, Germany}

\author{Bassano Vacchini}

\affiliation{Dipartimento di Fisica dell'Università di Milano and
             INFN Sezione di Milano, Via Celoria 16, I-20133 Milano, Italy}

\date{\today}

\begin{abstract}
We construct a large class of non-Markovian master equations that
describe the dynamics of open quantum systems featuring strong
memory effects, which relies on a quantum generalization of the
concept of classical semi-Markov processes. General conditions for
the complete positivity of the corresponding quantum dynamical maps
are formulated. The resulting non-Markovian quantum processes allow
the treatment of a variety of physical systems, as is illustrated
by means of various examples and applications, including quantum
optical systems and models of quantum transport.
\end{abstract}

\pacs{03.65.Yz,42.50.Lc,02.50.Ga,03.65.Ta}

\maketitle

The analysis of the time evolution of open systems plays a central
role in many applications of modern quantum theory, including
quantum information science, quantum transport theory, quantum
thermodynamics, and quantum process tomography and control (see
e.g. \cite{general}). The state of an open quantum system that is
coupled to the degrees of freedom of its surroundings is
represented by a time-dependent density matrix $\rho(t)$. In the
Markovian regime the dynamics is governed by a master equation of
the relatively simple form
\begin{equation} \label{LINDBLAD} \frac{d}{dt}\rho(t) =
   {\mathcal{L}}\rho(t),
\end{equation}
where ${\mathcal{L}}$ is a time-independent generator with the
famous Gorini-Kossakowski-Sudarshan-Lindblad structure
\cite{GORINI-LINDBLAD}
\begin{equation} \label{L-STRUCTURE} {\mathcal{L}}\rho = -i[H,\rho] +
   \sum_{\alpha} \left[ A_{\alpha} \rho A^{\dagger}_{\alpha} -
      \frac{1}{2} \left\{A^{\dagger}_{\alpha}A_{\alpha},\rho\right\}
   \right].
 \end{equation}
The Hamiltonian $H$ describes the coherent part of the time
evolution and the $A_{\alpha}$ are certain operators representing
the various decay modes. The solution of Eq.~(\ref{LINDBLAD}) can
be written in terms of a linear map $V(t)=\exp({\mathcal{L}}t)$
that transforms the initial state $\rho(0)$ into the state
$\rho(t)=V(t)\rho(0)$ at time $t$. The physical interpretation of
this map $V(t)$ requires that it preserves the trace and the
positivity of the density matrix $\rho(t)$. According to general
physical principles $V(t)$ must be a completely positive (CP) map
\cite{KRAUS,TheWork}. Hence, $V(t)$ represents a CP dynamical
semigroup known as quantum Markov process, whose generator has
been proven \cite{GORINI-LINDBLAD} to be of the
form~\eqref{L-STRUCTURE}.

The quantum dynamics given by Eq.~\eqref{L-STRUCTURE} has a
clearcut connection to a classical Markov process for the case in
which one has a closed system of equations for the populations
$P_n (t)=\langle n|\rho (t)|n\rangle$ in a fixed orthonormal basis
$\{|n\rangle\}$ of the open system's Hilbert space, typically the
energy eigenbasis. In fact, in this case one recovers the Pauli
master equation,
\begin{equation}
    \label{eq:5}
       \frac{d}{dt}
P_n (t)
=   \sum_m
 \left[ \Gamma_{nm}
P_m (t)
 - \Gamma_{mn}P_n (t) \right],
\end{equation}
which describes a classical Markovian jump process with transition
rates $\Gamma_{mn}$, justifying the notion of a quantum Markov
process.

The most important physical assumption which underlies the master
equation (\ref{LINDBLAD}) is the validity of the Markov
approximation of short environmental correlation times. If this
approximation is violated non-Markovian dynamics emerges which is
characterized by pronounced memory effects, finite revival times
and non-exponential relaxation and decoherence. These effects can
result from long-range correlation functions, from correlations
and entanglement in the initial state, as well as from the
neglection of extra degrees of freedom affecting the dynamics
\cite{BUDINI,nmgroup}. As a consequence the theoretical treatment
of non-Markovian quantum dynamics is generally extremely
demanding. A widely used non-Markovian generalization of
Eq.~(\ref{LINDBLAD}) is given by the integrodifferential equation
\begin{equation}
\label{NOnon-MarkovianARKOV}
 \frac{d}{dt}\rho(t) = \int_0^t d\tau \, {\mathcal{K}}(\tau) \rho(t-\tau).
\end{equation}
In this equation one takes into account quantum memory effects
through the introduction of the memory kernel
${\mathcal{K}}(\tau)$ which means that the rate of change of the
state $\rho(t)$ at time $t$ depends on the states $\rho(t-\tau)$
at previous times $t-\tau$. Equations of the form
(\ref{NOnon-MarkovianARKOV}) arise, for instance, by employing the
standard Nakajima-Zwanzig projection operator technique
\cite{NAKAJIMA-ZWANZIG}. Obviously, the Markovian master equation
(\ref{LINDBLAD}) is obtained if the memory kernel is taken to be
proportional to a $\delta$-function,
${\mathcal{K}}(\tau)=2\delta(\tau){\mathcal{L}}$.

In order to be physically acceptable the superoperator
$\mathcal{K} (\tau)$ appearing in Eq.~\eqref{NOnon-MarkovianARKOV}
must grant the CP of the resulting quantum dynamical map $V(t)$.
This is a very stringent requirement and, in fact, the general
structural characterization of physically admissible memory
kernels is an unsolved problem of central importance in the field
of non-Markovian quantum dynamics \cite{BUDINI,Daffer-Lidar}. It
has been realized recently that even the most simple and natural
choices for the memory kernel can lead to unphysical results
\cite{BARNETT,BUDINI}. To improve this situation we will construct
a class of non-Markovian quantum master equations that arises
naturally as a quantum mechanical generalization of classical
semi-Markov processes \cite{Feller}. The approach proposed here
leads to important physical insights guiding the phenomenological
determination of the memory kernel, and, at the same time, enables
a compact formulation of sufficient conditions that guarantee the
existence and the CP of the quantum dynamical map. Moreover, for a
specific class of processes one can formulate CP conditions which
are not only sufficient but also necessary.

We consider memory kernels with the general structure
\begin{eqnarray}
   \label{eq:1}
   \mathcal{K} (\tau)\rho&=& -i\left[H(\tau),\rho\right]
 - \frac 12 \sum_{\alpha}
\left\{A^{\dagger}_{\alpha}(\tau)A_{\alpha}(\tau),\rho\right\}
\nonumber \\
&~&
+\sum_{\alpha} A_{\alpha}(\tau)\rho A^{\dagger}_{\alpha}(\tau),
\end{eqnarray}
that is to say of the form given by Eq.~\eqref{L-STRUCTURE} apart
from the time dependence of the considered operators. As
previously done in the Markovian case let us consider the
situation in which the populations obey a closed system of
equations of motion, which then takes the form
\begin{eqnarray} \label{eq:4}
 \frac{d}{dt} P_n (t) &=& \int_0^t d\tau \sum_m
 \Big[ W_{nm} (\tau) P_m (t-\tau)
 \nonumber \\
 &~&\qquad \qquad \qquad - W_{mn} (\tau)P_n (t-\tau) \Big],
\end{eqnarray}
where $W_{nm}(\tau)=\sum_\alpha |\langle n
|A_{\alpha}(\tau)|m\rangle|^2$. This is the master equation for a
general type of classical non-Markovian processes known as
semi-Markov processes \cite{Feller}. Thus, whenever the
populations obey closed equations, Eq.~\eqref{L-STRUCTURE} yields
the classical Markovian master equation \eqref{eq:5}, while
Eq.~\eqref{NOnon-MarkovianARKOV} with the kernel \eqref{eq:1}
leads under the same conditions to the generalized master equation
(\ref{eq:4}) for a classical semi-Markov process. This justifies
on the same footing as before the name quantum semi-Markov
process.

To clarify the physical content of Eq.~\eqref{eq:4} let us
consider as an example the situation in which the kernel functions
$W_{nm} (t)$ factorize as $W_{nm} (t)=\pi_{nm} k_m (t)$, where
$\pi_{nm}\geq 0$ and $\sum_n \pi_{nm}=1$. The corresponding
process can then be interpreted as describing a particle moving on
a lattice with sites labelled by $n$, where the $\pi_{nm}$ are the
probabilities for jumps from site $m$ to site $n$. Jumps out of a
given site $n$ take place after a certain waiting time $t$ that
follows the waiting time distribution $f_{n} (t)$. The
characteristic feature of semi-Markov processes is the fact that,
by contrast to the Markovian case,  $f_{n} (t)$ need not be an
exponential function, but can be any probability distribution,
thus giving rise to memory effects. These waiting time
distributions are uniquely determined by the functions $k_n(t)$
according to the relation \cite{GILLESPIE}
\begin{equation}
\label{eq:2}
   f_{n} (t) =\int_0^t d\tau\, k_n (\tau)g_{n}(t-\tau)\equiv (k_n \ast
g_{n})(t),
\end{equation}
where the function
\begin{equation}
   \label{eq:3}
   g_{n} (t)=1-\int_0^{t} d\tau \, f_{n} (\tau)
\end{equation}
denotes the probability not to have left site $n$ by time $t$, the
so-called survival probability, and $\ast$ is the usual
convolution product. The generalized master equation~\eqref{eq:4}
therefore provides a physically acceptable time evolution for the
populations $P_n (t)$, granting in particular their positivity,
provided the functions $k_n(t)$ allow an interpretation in terms
of waiting time distributions \cite{GILLESPIE,Esposito2008a}.

However, these classical conditions are clearly not enough to
ensure the existence of a well-defined dynamics in the quantum
case, and a general characterization at the quantum level can
hardly be achieved. Therefore our next goal is the formulation of
sufficient conditions that guarantee the CP of the dynamical map
$V(t)$ corresponding to the non-Markovian master equation defined
by Eqs.~(\ref{NOnon-MarkovianARKOV}) and \eqref{eq:1}, no longer
assuming that closed equations for the populations exist. This map
is defined by
\begin{equation} \label{VT}
 \frac{d}{dt}V(t) = \int_0^t d\tau \, {\mathcal{K}}(\tau) V(t-\tau),
\end{equation}
together with the initial condition $V(0)=I$, with $I$
the identity map.
We now employ ideas recently
formulated in Ref.~\cite{KOSSAKOWSKI}, decomposing the memory
kernel as ${\mathcal{K}}(\tau) = B(\tau) + C(\tau)$, where
$B(\tau)$ is the CP map defined by
\begin{equation}
   \label{eq:7}
   B(\tau)\rho = \sum_{\alpha} A_{\alpha}(\tau)\rho A^{\dagger}_{\alpha}(\tau),
\end{equation}
and $C(\tau)$ is given by the first line of \eqref{eq:1}. We
further introduce the map $V_0(t)$ as the solution of the equation
\begin{equation} \label{R-EQ}
 \frac{d}{dt}V_0(t) = \int_0^t d\tau\, C(\tau) V_0(t-\tau),
\end{equation}
with the initial condition $V_0(0)=I$. Considering the Laplace
transforms of Eqs.~\eqref{VT} and \eqref{R-EQ} one obtains a
resolvent-like identity for the dynamical map leading in the time
domain to the equation
\begin{equation} \label{DYSON}
 V(t) = V_0(t) + (V_0 \ast B \ast V)(t).
\end{equation}
Regarding formally the superoperator $B(\tau)$ as a perturbation
and iterating Eq.~(\ref{DYSON}) one finds that the full dynamical
map $V(t)$ can be represented as a series,
\begin{eqnarray} \label{V-REP}
 V(t) &=& V_0(t) + (V_0\ast B\ast V_0)(t) \nonumber \\
 &~& + (V_0\ast B\ast V_0\ast B \ast V_0)(t) + \ldots.
\end{eqnarray}
Due to the fact that the set of CP maps is closed under addition
and convolution, we can conclude from Eq.~\eqref{V-REP} that $V
(t)$ is CP if $V_0(t)$ is CP. To bring  this condition into an
explicit form let us assume that the Hermitian operators $H(\tau)$
and $\sum_{\alpha}A^{\dagger}_{\alpha}(\tau)A_{\alpha}(\tau)$ are
diagonal in the time-independent orthonormal basis
$\{|n\rangle\}$, that is $H(\tau) = \sum_n \varepsilon_n(\tau)
|n\rangle\langle n|$ and
\begin{equation}
   \label{eq:9}
   \sum_{\alpha} A^{\dagger}_{\alpha}(\tau)A_{\alpha}(\tau)
 = \sum_n k_n(\tau) |n\rangle\langle n|.
\end{equation}
Then we can solve Eq.~(\ref{R-EQ}) to obtain
\begin{equation} \label{R-REP}
 V_0(t)\rho (0)  = \sum_{nm} g_{nm}(t)
 |n\rangle\langle n|\rho (0) |m\rangle\langle m|,
\end{equation}
where the functions $g_{nm}(t)$ are the solutions of
\begin{equation} \label{Gnon-Markovian}
 \dot{g}_{nm}(t) = -\int_0^t d\tau
 \left[ z_n(\tau)+z_m^{\ast}(\tau) \right] g_{nm}(t-\tau),
\end{equation}
corresponding to the initial conditions $g_{nm}(0)=1$, and
$z_n(\tau)=\frac{1}{2}k_n(\tau)+i\varepsilon_n(\tau)$. To prove
Eq.~(\ref{R-REP}) one first shows that
$C(\tau)\left(|n\rangle\langle
m|\right)=-[z_n(\tau)+z_m^{\ast}(\tau)]|n\rangle\langle m|$. Using
this relation one easily demonstrates that the expression
(\ref{R-REP}) indeed represents the desired solution of
Eq.~(\ref{R-EQ}). It is important to notice that the functions
$g_{nn} (t)$ do actually coincide with the survival probabilities
$g_{n} (t)$ introduced by Eq.~\eqref{eq:3}.

Employing the Kraus representation \cite{KRAUS} we see that the
map $V_0(t)$ given by Eq.~(\ref{R-REP}) is CP if and only if the
matrix with elements $g_{nm}(t)$ is positive,
\begin{equation} \label{COND-1}
 G(t) = (g_{nm}(t)) \geq 0.
\end{equation}
Hence, we arrive at a sufficient condition for CP: The quantum
dynamical map $V(t)$ corresponding to the non-Markovian master
equation (\ref{NOnon-MarkovianARKOV}) with the memory
kernel~\eqref{eq:1} is CP if the condition (\ref{COND-1}) is
fulfilled. A necessary condition for \eqref{COND-1} to hold is the
positivity of the diagonal elements of $G (t)$, which are given by
the survival probabilities $g_n(t)=g_{nn}(t)$. This necessary
condition in turn implies the positivity of the functions $f_{n}
(t)$ according to Eq.~\eqref{eq:2}, which can then be interpreted
as true waiting time distributions. The positivity of the matrix
$G(t)$ therefore represents a natural quantum generalization of
the classical conditions.

We illustrate the result (\ref{COND-1}) with the help of several
examples, which all fall into the class of quantum semi-Markov
processes introduced by means of Eqs.~\eqref{NOnon-MarkovianARKOV}
and \eqref{eq:1}. A prototypical system showing strong
non-Markovian behavior is a two-level atom interacting with a
damped field mode described by the memory kernel
\begin{eqnarray} \label{JC}
 {\mathcal{K}}(\tau)\rho  &=&
 -i\varepsilon(\tau)[\sigma_+\sigma_-,\rho ] \nonumber \\
 &~& +k(\tau) \left[ \sigma_-\rho \sigma_+
 - \frac{1}{2} \left\{ \sigma_+\sigma_-,\rho \right\} \right].
\end{eqnarray}
Excited and ground state are denoted by $|+\rangle$ and
$|-\rangle$, respectively, and $\sigma_{\pm}$ are the
corresponding raising and lowering operators. The index $n$ thus
takes on the two values $n=\pm$. For a positive function $k(\tau)$
the memory kernel (\ref{JC}) is of the form specified above with
$k_+(\tau)=k(\tau)$, $k_-(\tau)=0$,
$\varepsilon_+(\tau)=\varepsilon(\tau)$, and
$\varepsilon_-(\tau)=0$. Hence, the matrix $G(t)$ takes the form
\begin{equation}
 G(t) = \left( \begin{array}{cc}
 g_{++}(t) & g_{+-}(t) \\
 g_{+-}^{\ast}(t) & 1
 \end{array} \right),
\end{equation}
with $g_{++}(t)$ and $g_{+-}(t)$ determined by
Eq.~(\ref{Gnon-Markovian}). Thus we see that the condition
(\ref{COND-1}) for CP is equivalent to $g_{++}(t) \geq
|g_{+-}(t)|^2$. The master equation corresponding to the memory
kernel (\ref{JC}) can be solved analytically. One then finds that
for this case the condition \eqref{COND-1} is not only sufficient
but also necessary for CP.

A further very instructive example involving an infinite
dimensional Hilbert space is the model of a quantum oscillator
with non-Markovian damping studied in Ref.~\cite{BARNETT}. The
memory kernel for this model reads
\begin{equation}\label{aa}
 {\mathcal{K}}(\tau)\rho  = k(\tau) \left[
 a\rho a^{\dagger}-\frac{1}{2}\left\{a^{\dagger}a,\rho \right\} \right],
\end{equation}
where $k(\tau)=\kappa\exp(-\gamma\tau)$ and $a^{\dagger}$, $a$ are
the raising and lowering operators of the oscillator. This kernel
is again of the form~\eqref{eq:1} with a single Lindblad operator
$A (\tau)=\sqrt{k (\tau)}a$. Here, the basis states $|n\rangle$
are the number states of the oscillator, $k_n(\tau)=nk(\tau)$ and
$\varepsilon_n(\tau)=0$. Solving Eq.~(\ref{Gnon-Markovian}) by
means of a Laplace transformation, we find
\[
 g_{nn}(t) = e^{-\gamma t/2} \left[ \cosh(d_nt/2) + \frac{\gamma}{d_n}
 \sinh(d_nt/2)\right],
\]
where $d_n=\sqrt{(\gamma/2)^2-n\kappa}$. For the necessary
condition $g_{nn}(t)\geq 0$ to hold $d_n$ must be real. This shows
that condition (\ref{COND-1}) is certainly violated if $4n\kappa >
\gamma^2$. Because $n$ can be arbitrary large we conclude that
condition (\ref{COND-1}) is never fulfilled. The interesting
aspect of this example is the fact that the non-Markovian master
equation indeed violates not only CP but also positivity.  This
fact has been demonstrated in \cite{BARNETT} and clearly shows
again the relevance of our CP conditions.

Many further physical systems lead to a generalized master
equation of the form introduced here if one applies the
Nakajima-Zwanzig projection operator technique, such as the
tight-binding quantum diffusion model studied in \cite{GASPARD},
and the quantum transport model introduced in \cite{DIFBAL}, which
leads to a memory kernel of the form
\begin{equation}
 {\mathcal{K}}(\tau)\rho  = k(\tau)
 \left[ \frac{1}{2}T \rho  T^{\dagger} + \frac{1}{2}T^{\dagger} \rho  T
 - \rho  \right].
\end{equation}
This kernel describes the motion of an excitation in a modular
system consisting of weakly coupled subunits labelled by the index
$n$, where $T=\sum_n|n+1\rangle\langle n|$ represents the
corresponding translation operator. The model features strong
non-Markovian behavior and a transition from diffusive to
ballistic quantum transport. The memory kernel
${\mathcal{K}}(\tau)$ is obviously of the form introduced above.
The Hamiltonian contribution vanishes, $H(\tau)=0$, and all kernel
functions are equal to each other, $k_n(\tau)=k(\tau)$, which
corresponds to the special case treated in
Refs.~\cite{KOSSAKOWSKI,BUDINI} with a loss term proportional to
the identity operator. Equation (\ref{Gnon-Markovian}) shows that
also all matrix elements of $G(t)$ are equal, $g_{nm}(t)=g(t)$,
and, hence, condition (\ref{COND-1}) reduces to the condition
$g(t)\geq 0$. Clearly this condition leads to important
restrictions on the form of the kernel function $k(\tau)$ which is
determined by the correlation function of the microscopic model.

As our final example we discuss memory kernels of the following
general structure,
\begin{eqnarray} \label{QSM}
 {\mathcal{K}}(\tau)\rho  &=&
 -i\left[H(\tau),\rho \right] -\frac 12 \sum_n
 k_n(\tau) \left\{|n\rangle\langle n|,\rho \right\},
\nonumber
 \\
 &~&+
\sum_{nm} \pi_{nm} k_m(\tau) |n\rangle\langle m|\rho |m\rangle\langle n|
\end{eqnarray}
of which \eqref{JC} provides an example. For this memory kernel
the coherences of the density matrix, i.~e. the off-diagonal
elements $\rho_{nm}(t)=\langle n|\rho(t)|m\rangle$, $n\neq m$, are
simply given by $\rho_{nm}(t)=\rho_{nm}(0)g_{nm}(t)$. On the other
hand, the diagonals of the density matrix, i.~e. the populations
$P_n(t)$ obey a closed transport equation as in~\eqref{eq:4}. It
is remarkable that in this case one can go one step further to
derive a condition for the CP which is not only sufficient but
also necessary. To this end one writes the quantum dynamical map
$V(t)$ corresponding to the non-Markovian quantum master equation
(\ref{NOnon-MarkovianARKOV}) with the memory kernel (\ref{QSM}) in
terms of the functions $g_{nm}(t)$ and of the conditional
transition probabilities $T_{nm}(t)$ obeying the classical master
equation~\eqref{eq:4}. The quantity $T_{nm}(t)$ represents the
probability that the particle is at site $n$ at time $t$ given
that it started at site $m$ at time $t=0$. With the help of the
resulting expression for the map $V(t)$ we then find the following
result. Given a classical semi-Markov process, the quantum
dynamical map $V(t)$ is CP if and only if the condition
\begin{equation} \label{COND-2}
 \tilde{G}(t) = (\tilde{g}_{nm}(t)) \geq 0
\end{equation}
is satisfied. Here, the off-diagonal elements of the matrix
$\tilde{G}(t)$ coincide with those of $G(t)$, while the diagonals
of $\tilde{G}(t)$ are given by the conditional transition
probabilities, $\tilde{g}_{nn}(t)=T_{nn}(t)$. Note that the
probabilities $T_{nn}(t)$ are in fact in general greater than the
corresponding survival probabilities $g_{nn}(t)$, since the system
can be in state $n$ at time $t$ both because it has not left it
and because it has come back to the initial state.
Eq.~\eqref{COND-2} thus provides a complete characterization of
the CP of the class of quantum semi-Markov processes given
by~\eqref{QSM}.

Building on an analogy with classical semi-Markov processes we
have constructed a large class of non-Markovian master equations
with memory kernel and formulated sufficient conditions for the CP
of the resulting quantum dynamical map. The latter impose strong
restrictions on the structure of physically acceptable
non-Markovian quantum master equations, which are particularly
useful in phenomenological approaches.  For a specific class of
quantum semi-Markov processes necessary and sufficient conditions
for CP have also been formulated. Important further developments
of the theory should include the case of temporarily negative
kernel functions and effects from correlations and entanglement in
the initial state.

\begin{acknowledgments}
One of us (HPB) gratefully acknowledges a Fellowship of the
Hanse-Wissenschaftskolleg, Delmenhorst.
\end{acknowledgments}

\end{document}